\documentclass[10pt,a4paper]{article}
\usepackage{hyperref}
\usepackage{amsmath}
\usepackage[psamsfonts]{amssymb }
\usepackage{mathrsfs}

\usepackage{dsfont}

\begin{document}

\title{Towards the correct prescription of the Problem of divergence of Propagator of Siegal-Zwiebach action from String theory in Fierz-Pauli Gauge in the small mass limit}

\author{Vipul Kumar Pandey${}$\footnote {vipulvaranasi@gmail.com} \\\\
${}$Department of Physics,
Chandigarh University\\
Mohali -- 140413, India\\\\
}
\date{}
\maketitle

\begin{abstract}
In the present work we will give an explicit solution the problem of divergence of propagator of gauge-invariant Siegel-Zwiebach (SZ) action in Fierz-Pauli (FP) gauge by connecting it's Green's functions to that of Transverse-Traceless (TT) gauge using improved finite-field-dependent BRST (FFBRST) method.
\end{abstract}

\section{Introduction} 
Since Dirac's initial venture to canonically quantization the Einstein’s theory of General relativity\cite{Dirac:1958sc, PDIR} and followed by Feynman's \cite{RPF} and Weinberg's \cite{SW1,SW2,SW3}, numerous attempts for constructing 
 a quantum mechanical formulation of the spin-two particle have been made. In spite of various efforts made, construction of a fully satisfactory theory of quantum gravity is still very much illusive. The major problems with the Einstein's theory of gravity are its nonlinear and non-renormalization nature along with dynamical nature of space-time metric.
Besides that, there are other problems arising from experimental observations revealing an accelerating expanding universe containing unknown forms of matter and energy. In this condition, inventive modifications of Einstein’s theory of gravity and their currently established theoretical frameworks are desirable and definitely deserves the physicist's attentiveness \cite{Reyes:2022dil, Reyes:2022mvm, Pfeifer:2023cgd}. We would like to address such possible modification, like massive gravity, which allows for the possibility for existence of non-null small mass for the graviton.However, the major problem with a massive graviton hypothesis is the immediate breaking of gauge symmetry, leading to a discontinuity of theoretical predictions done by van Dam, Veltman and Zakharov \cite{vanDam:1970vg, Zakharov:1970cc} which has also been revisited recently in the context of higher derivatives theories \cite{Myung:2017zsa}. It has also been observed that the number of degrees of freedom changes drastically in the massless limit. Reviews of massive gravity from start to present status, including the Vainshtein solution\cite{Vainshtein:1972sx}
to the van Dam-Veltman-Zakharov discontinuity, can be found in \cite{KH,CDR}. Another existing proposal in the literature is the aspiring idea of obtaining a consistent elucidation of quantum gravity from point of view of string theory with the possible reward of its unification with other fundamental forces of nature.  
Those two programs can be surprisingly related to each other. As a reference, we know that higher-spin fields actions can be obtained from string field theory \cite{WSBZ, Bengtsson:1986ys, Sagnotti:2003qa, Asano:2012qn, Asano:2013rka, Lee:2017utr, HPTL}. It is also important to mention here that the study of small mass spin-two particle may give a better understanding of open strings in the high energy limit\cite{Gross:1987ar}.

 In the present work, we discuss the linearized massive gravity which is described by the embedding of the spin-two Fierz-Pauli model \cite{MFWP} into greater space defined by the gauge-invariant Siegel-Zwiebach (SZ) action \cite{WSBZ} from bosonic string theory in the critical dimension.
In \cite{WSBZ}, Siegel and Zwiebach derived a full gauge-invariant action defining the bosonic string using a second-quantized BRST-invariant formulation.  Auxiliary fields corresponding to higher-spin modes present in Lagrangian functions and contained in the string are obtained by expansions into component fields from the main functional field. More precisely, Siegel and Zwiebach have produced a Lagrangian density characterizing a massive spin-two field interacting with two extra vector and scalar fields \cite{WSBZ}, by expanding the open string action to fourth mass level. This enormous rank two symmetric tensor, or spin-two field, can be understood as a massive graviton, whose characteristics will be discussed in more detail later on.  Because of its origin, the heavy graviton must to live inside the bosonic string critical dimension in order to experience proper gauge invariance.
Recently, Park and Lee \cite{HPTL} have proposed the use of the SZ model in a transverse-traceless gauge to study small mass gravitons from string theory.  
Additionally to substantial calculation simplifications, a transverse-traceless gauge has the advantage of propagating only physical degrees of freedom \cite{MTW, Carroll:2004st}. It has been observed in recent literature that, in the traditional approach to linearized massless gravity, transverse and traceless propagator is not achievable by usual Faddeev-Popov procedure \cite{BFM}.

Park and Lee have achieved a transverse traceless finite propagator for the massive graviton in the bosonic string context, starting from the corresponding classical Lagrangian and utilizing intriguing heuristic arguments that include comparisons with the massive Proca model and performing direct substitution of subsidiary gauge conditions back into the SZ Lagrangian itself\cite{HPTL}.  Although this may not appear to be the case at the moment, quantization of gauge-invariant systems can be difficult and even misleading, as demonstrated by several cases in quantum field theory. Here, we continue from our earlier work\cite{PT} where we took a different approach and aimed to formalize and defend the concept of Park and Lee inside a stricter functional quantization framework. More specifically, we have employed the two versions of the BRST quantization technique, which entail finite-field-dependent and infinitesimal transformations. We have ensured unitarity by casting the SZ action into a BRST invariant form and computing the appropriate Green functions generating functional within a covariant method in a wider class of covariant gauges by including the required ghost and auxiliary fields\cite{PT}. 
In our earlier work, we have functionally quantized the SZ model in a full-controlled consistent BRST-invariant manner\cite{PT}. The gauge-fixed SZ is demonstrated to be left invariant under normal BRST transformations, including a matching Jacobian to account for the functional integration measure change under finite-field-dependent BRST (FFBRST) transformations \cite{JM1}. It is demonstrated that an FFBRST transformation connects the unitary and generalized Lorenz type gauges precisely via the use of that nontrivial Jacobian. In this approach, we have also validated the consistency of the small mass graviton propagator that is found in the transverse-traceless case as well as in the entire class of covariant gauges that are explored. In the present work we have explicitly shown how the Green's functions correspondings to two set of gauges can be connencted through the improved FFBRST method developed in \cite{BJ} and subsequently discussed in \cite{JAM1,JAM2,JM,UM}. This leads to a finite value of propoagator in the small mass limit, hence removing the divergence in the SZ action in FP gauge condition. 
       
The structure of the current work is as follows.The Siegel-Zwiebach action, which describes a massive rank-two tensor interacting with a scalar and vector fields that enjoy gauge invariance in the string critical dimension, is introduced together with our terminology and conventions in the next section. As a special case, the enormous Fierz-Pauli action is found to follow the unitary gauge. In section {\bf 3}, we will discuss about a two-parameter generalized Lorenz type gauge-fixing also called transverse-traceless gauge-fixing for the SZ action in a functional quantization framework. The transverse-traceless gauge attainability is explicitly shown and can be obtained at quantum level for specific limits of the gauge parameters. The small mass graviton propagator in the generalized Lorenz gauge is obtained and verified to have a fine well behaved massless transverse-traceless limit.

We shall talk about a two-parameter transverse-traceless gauge-fixing, also known as a generalized Lorenz type gauge-fixing, for the SZ action in a functional quantization framework in section {\bf 3}. A fine well behaved massless transverse-traceless limit is obtained and verified for the small mass graviton propagator in the generalized Lorenz gauge.

In section {\bf 4}, we give a brief summary of the Improved FFBRST formalism and proceed to show that the Green's functions corresponding to the unitary and generalized Lorenz type gauges can be connected by a IFFBRST transformation providing a formal proof of their equivalence of propagators in two gauges and hence showing the finiteness of SZ propagator in FP gauge (unitary gauge) in the small mass graviton limit leading to a formal prescription of vDVW discontinuity. We end in section {\bf 5} with some concluding remarks.

\section{Gauge-invariant massive Siegel-Zwiebach action in Fierz-Pauli Gauge}
In string theory, massive spin-two fields arise in the spectrum of open strings as part of a corresponding rank-two tensor multiplet. Siegel and Zwiebach obtained\cite{WSBZ}, a BRST-invariant formulation for the free bosonic string, expanding its functionals in terms of component fields in $D=26$ space-time dimensional gauge invariant massive spin-two action describing a massive symmetric tensor field $h_{\mu\nu}$, with trace denoted by $h \equiv h^{\mu}_{\;\mu}$, coupled to an auxiliary vector field $B^\mu$ and a scalar field $\theta$ as \cite{PT}
\begin{eqnarray}
&&{S}_{SZ} = \int d^{26}x\, \frac{1}{2}\left\{ \frac{1}{2}h_{\mu\nu}{(\Box - m^2)}h^{\mu\nu} + B_\mu{(\Box - m^2)} B^\mu \right. \nonumber \\&& \left.- \theta{(\Box - m^2)} \theta + {\left(\partial^\nu h_{\mu\nu} + \partial_\mu \theta - m B_\mu\right)^2} \right. \nonumber \\&& \left.+ {\left(\frac{mh}{4} + \frac{3m\theta}{2} + \partial_\mu B^\mu\right)^2} \right\}
\label{LSZ}
\end{eqnarray}
It can be easily verified that SZ action above (\ref{LSZ}) is invariant under the local gauge transformations defined by 27 arbitrary space-time dependent parameters $\epsilon$ and $\epsilon^\mu$.({The metric of flat Minkowski space-time is denoted by $\eta^{\mu\nu}$ with signature convention $(-1,1,\dots,1)$ and Greek indexes running through $0,1,\dots,25$.})\cite{PT}
\begin{equation}
\begin{split}
\delta h_{\mu\nu} &= \partial_\mu \epsilon_{\nu} + \partial_\nu \epsilon_{\mu} - \frac{1}{2}m\eta_{\mu\nu}\epsilon\,,\\
\delta B_{\mu} &= \partial_\mu \epsilon + m \epsilon_\mu\,,\\
\delta \theta &= - \partial_\mu \epsilon^\mu + \frac{3}{2}m\epsilon\,,
\end{split}
\label{gs}
\end{equation}
It works exactly for the critical dimension $D=26$ which can be understood by going back to gauge symmetry (\ref{gs}) and its  origin corresponding to the nilpotency of the BRST charge in open bosonic string theory, which is a necessary condition for the absence of anomaly at quantum level.  
 
Now, we will substitute a gauge condition 
\begin{eqnarray}
B^\mu = 0, \quad \theta = -\frac{h}{2}\,,
\label{FPG}
\end{eqnarray}
into (\ref{LSZ}) which will lead to a new action 
\begin{eqnarray}
&&{S}_{FP} = \int d^{26}x\, \left\{ \frac{1}{4}h_{\mu\nu}{(\Box - m^2)}h^{\mu\nu}
+ \frac{1}{2} \partial^\nu h_{\mu\nu} \left( \partial^\rho h^\mu_{\;\rho}-\partial^\mu h \right)
\right. \nonumber \\&& \left. - \frac{1}{4} h{(\Box - m^2)} h \right\} \,,
\label{SFP}
\end{eqnarray}
It is called a non-gauge-invariant massive Fierz-Pauli (FP) action as it was first proposed by Fierz and Pauli\cite{MFWP}.  Here, $B^\mu$ and $\theta$ acts as Stueckelberg fields \cite{Stueckelberg:1957zz} reassuring gauge-invariance for the massive case.  Also, the massless case in (\ref{SFP}) will give linearized Einstein-Hilbert action having invariance under\cite{PT}
\begin{equation}
\delta h_{\mu\nu} = \partial_\mu \epsilon_{\nu} + \partial_\nu \epsilon_{\mu} 
\,.
\end{equation} 
This suggests that (\ref{FPG}) can be a preferable gauge-fixing condition for (\ref{LSZ}), playing the role of a unitary gauge condition relating corresponding first and second-class systems \cite{Amorim:1999xr}. 

Using the BRST quantization framework \cite{Becchi:1974xu, Becchi:1974md, Tyutin:1975qk} the nilpotent BRST transformation for this system can be written as\cite{PT}
\begin{equation}
\begin{array}{c}
s h_{\mu\nu} = \partial_\mu c_{\nu} + \partial_\nu c_{\mu} - \displaystyle\frac{m}{2}\eta_{\mu\nu}c\,,\\[10pt]
s B_{\mu} = \partial_\mu c + m c_\mu\,,\quad
s \theta = - \partial_\mu c^\mu + \displaystyle\frac{3}{2}m c \,,\\[10pt]
s c_{\mu} = 0 \,,\quad
s \bar{c}_{\mu} = b_\mu\,,\quad
\quad s b_\mu = 0\,,
 \\[10pt] 
s c = 0 \,,\quad 
s \bar{c} = b \,,\quad 
s b = 0\,, 
\end{array}
\label{BRST}
\end{equation}
Where $(c, \bar{c})$ and $(c^\mu,\bar{c}_\mu)$ are anticommuting ghost-antighost fields and $(b,b^\mu)$ are the Nakanishi-Lautrup auxiliary fields.

Now using (\ref{BRST}) and (\ref{FPG}) the gauge-fixing and ghosts terms can be written as
\begin{equation}
S_{ug} = \int d^{26}x\, 
\left[b_\mu B^\mu + b\theta +\frac{bh}{2}
-{5m}\bar{c}c+\bar{c}_\mu \partial^\mu c + m \bar{c}_\mu c^\mu 
\right]
\,,
\end{equation}
Now the generating functional for total effective action in FP gauge is written as\cite{PT}
\begin{equation}
Z_u = \int d\mu \, e^{i \displaystyle \left\{ {S}_{SZ} + S_{ug} \right\}}
\label{Z0}
\end{equation}
with integration measure defined as\cite{PT}
\begin{equation}
d\mu \equiv [dh_{\mu\nu}][dB_\mu][d\theta][dc][dc_\mu]
[d\bar{c}][d\bar{c}_\mu][db][db_\mu]
\label{IM}
\end{equation}
Now, it can be easily verified that generating functional (\ref{Z0}) is invariant under BRST transformations constructed in (\ref{BRST}) including a global infinitesimal anticommuting parameter.  Performing  the functional integration of (\ref{Z0}) wrt. fields $b$, $\theta$, $b^\mu$ and $B^\mu$ by calculationg, the massive spin two field propagator in the FP Gauge can be written as\cite{PT}
\begin{eqnarray}
&&G^{\alpha\beta \sigma\lambda} = \frac{i}{p^2 + m^2}\Bigg\{ \frac{2}{25}\left( \eta^{\alpha\beta} + \frac{p^\alpha p^\beta}{m^2}\right)\left( \eta^{\sigma\lambda} + \frac{p^\sigma p^\lambda}{m^2}\right) \nonumber\\&& - \left( \eta^{\alpha\sigma} + \frac{p^\alpha p^\sigma}{m^2}\right)\left( \eta^{\beta\lambda} + \frac{p^\beta p^\lambda}{m^2}\right) - \left( \eta^{\alpha\lambda} + \frac{p^\alpha p^\lambda}{m^2}\right)\nonumber\\&&\left( \eta^{\beta\sigma} + \frac{p^\beta p^\sigma}{m^2}\right)   \Bigg \}\,,
\label{PPFPL}
\end{eqnarray}
which enjoys following symmetry
\begin{equation}
G^{\alpha\beta \sigma\lambda} = G^{\beta\alpha \sigma\lambda} = G^{\alpha\beta \lambda\sigma}
= G^{\sigma\lambda \alpha\beta}
\,,
\end{equation}
describing the massive graviton. However, as can be immediately observed from (\ref{PPFPL}), the massless limit is not well-defined. Since our interested is around small mass gravitons, we may look for a more suitable gauge-fixing condition for the general action (\ref{LSZ}). It will be attempted in the next section.

\section{Gauge-invariant massive Siegel-Zwiebach action in Transverese-traceless Gauge}
We have observed above that, whilst compatible, the Fierz-Pauli gauge leads to a ill-defined propagator for the graviton in the massless limit. On the other hand, it can be shown \cite{HPTL} that a transverse-traceless (TT) gauge (generalization of the usual Lorenz condition in QED for the spin-two field  $h_{\mu\nu}$, along with a traceless condition) works better. TT gauge condition can be written as \cite{MTW}
\begin{equation}
\partial_\mu h^\mu_{\,\,\nu} = 0 \mbox{~~~and~~~} h=0 \,,
\label{TTgauge}
\end{equation}

The BRST invariant generating functional can be written as
\begin{equation}
Z = \int d\mu \, e^{i \displaystyle  S_{TT} }
\label{Z1}
\end{equation}
where gauge-fixed quantum action is defined as
\begin{equation}\label{Stt}
S_{TT} = {S}_{SZ} + S_{Tg}
\,,
\end{equation}
given by the sum of $S_{SZ}$ as in equation (\ref{LSZ}) and \cite{PT}
\begin{eqnarray}
&&S_{Tg} =
\int d^{26}x \left\{
\frac{\zeta b^2}{2m^2} + bh +\frac{\xi}{2}b^\mu b_\mu + b^\mu \partial_\nu h^\nu_{\,\mu}
+{\bar{c}}^\mu\Box c_\mu \right.\nonumber\\&&\left.+ \bar{c}^\mu \partial_\mu \partial_\nu c^\nu 
-\frac{m}{2} \bar{c}^\mu \partial_\mu c
+2 \bar{c} \partial_\mu c^\mu - 13 m \bar{c}c
\right\}
\,.
\end{eqnarray}
The  $d\mu$ in (\ref{Z1}) is the integration measure defined in (\ref{IM}).  

By performing a functional integration in (\ref{Z1}) over the field variables $b$, $b^\mu$, $\theta$ and $B_\mu$ and then integration by parts, we obtain the effective action as \cite{PT}
\begin{eqnarray}
&&S_{eff} = 
\int d^{26}x\,
\left\{
\frac{1}{2}
h_{\mu\nu}
{\cal O}^{\mu\nu\alpha\beta}
h_{\alpha\beta}
+{\bar{c}}^\mu\Box c_\mu + \bar{c}^\mu \partial_\mu \partial_\nu c^\nu 
\right.\nonumber\\&&\left.-\frac{m}{2} \bar{c}^\mu \partial_\mu c
+2 \bar{c} \partial_\mu c^\mu - 13 m \bar{c}c
\right\}
\end{eqnarray}
with
\begin{eqnarray}
&&{\cal O}^{\mu\nu\alpha\beta} \equiv
\frac{\Box-m^2}{4}
\left( \eta^{\mu\alpha}\eta^{\nu\beta}+\eta^{\mu\beta}\eta^{\nu\alpha} \right)
+\frac{12\Box^{-2}}{25}\left(\Box-m^2\right)
\nonumber\\&&\partial^\mu \partial^\nu \partial^\alpha \partial^\beta
+\frac{1}{4}\left[
\frac{1}{\xi}-\Box^{-1}(\Box-m^2)
\right]
\bigg(\eta^{\mu\alpha}\partial^\nu\partial^\beta+\eta^{\nu\alpha}\partial^\mu\partial^\beta\nonumber\\&&+\eta^{\mu\beta}\partial^\nu\
\partial^\alpha+\eta^{\nu\beta}\partial^\mu\partial^\alpha\bigg)
-\left[\frac{\Box-m^2}{25}+\frac{m^2}{\zeta}\right]
\eta^{\mu\nu}\eta^{\alpha\beta}
+\Box^{-1}\nonumber\\&&\left[\frac{\Box-m^2}{50}\right]
\left(
\eta^{\mu\nu}\partial^\alpha \partial^\beta+
\eta^{\alpha\beta}\partial^\mu \partial^\nu
\right)
\end{eqnarray}
from which we may compute the massive graviton propagator in momentum space as\cite{PT}
\begin{eqnarray}\label{G}
&&G_{\mu\nu\alpha\beta}^{\xi\zeta}=
-\frac{i\xi}{p^2}
\left[
\frac{p_\mu p_\nu p_\alpha p_\beta}{p^4}
+4\left(\eta_{\mu\alpha}-\frac{p_\mu p_\alpha}{p^2}\right)\frac{p_\beta p_\nu}{p^2}
\right.\nonumber\\&&\left.-\frac{2}{25}\left(\eta_{\mu\nu}-\frac{p_\mu p_\nu}{p^2}\right)\frac{p_\alpha p_\beta}{p^2}
+F^{\xi\zeta}(p^2,m^2)\left(\eta_{\mu\nu}-\frac{p_\mu p_\nu}{p^2}\right)\right.\nonumber\\&&\left.\left(\eta_{\alpha\beta}-\frac{p_\alpha p_\beta}{p^2}\right)
\right]
-\frac{2}{p^2+m^2}
\left[
\left(\eta_{\mu\alpha}-\frac{p_\mu p_\alpha}{p^2}\right)\left(\eta_{\nu\beta}-\frac{p_\nu p_\beta}{p^2}\right)
\right.\nonumber\\&&\left.-\frac{1}{25}\left(\eta_{\mu\nu}-\frac{p_\mu p_\nu}{p^2}\right)\left(\eta_{\alpha\beta}-\frac{p_\alpha p_\beta}{p^2}\right)
\right]
\end{eqnarray}
with
\begin{equation}
F^{\xi\zeta}(p^2,m^2)\equiv\frac{\xi}{1250p^2} \left[
\frac{p^2+m^2-50m^2\zeta^{-1}-50p^2\xi^{-1}}{p^2+m^2-50m^2\zeta^{-1}}
\right]
\,.
\end{equation}
It can be easily observed that, $G_{\mu\nu\alpha\beta}^{\xi\zeta}$ enjoys following symmetry properties
\begin{equation}
G^{\xi\zeta}_{\mu\nu \alpha\beta} = G^{\xi\zeta}_{\nu\mu \alpha\beta} = G^{\xi\zeta}_{\mu\nu \beta\alpha}
= G^{\xi\zeta}_{\alpha\beta \mu\nu}
\,.
\end{equation}
Different choices for the parameters $(\xi,\zeta)$, we lead to different set of gauges suitable for concrete calculations. So, the equation (\ref{G}) provides the propagator for a large class of covariant gauges with parameter $(\xi,\zeta)$. Specifically, in the limits $\xi\rightarrow0$, $\zeta\rightarrow 0$ we obtain following form of the propagator\cite{PT},
\begin{eqnarray}\label{G00}
&&G_{\mu\nu\alpha\beta}^{00}
=
-\frac{i}{p^2+m^2}
\left[
\left(\eta_{\mu\alpha}-\frac{p_\mu p_\alpha}{p^2}\right)\left(\eta_{\nu\beta}-\frac{p_\nu p_\beta}{p^2}\right)
\right.\nonumber\\&&\left.+
\left(\eta_{\mu\beta}-\frac{p_\mu p_\beta}{p^2}\right)\left(\eta_{\nu\alpha}-\frac{p_\nu p_\alpha}{p^2}\right)
\right.\nonumber\\&&\left.
-\frac{2}{25}\left(\eta_{\mu\nu}-\frac{p_\mu p_\nu}{p^2}\right)\left(\eta_{\alpha\beta}-\frac{p_\alpha p_\beta}{p^2}\right)
\right]
\end{eqnarray}
which is perfectly well defined in the massless limit.

\section{Finite Field BRST transformation}
The remarkable BRST symmetry was introduced through the groundbreaking works of Becchi, Rouet and Stora \cite{Becchi:1974xu, Becchi:1974md}
and Tyutin \cite{Tyutin:1975qk},
providing an requisite tool for the quantization of gauge invariant theories. In this symmetry transformation the original BRST parameter is infinitesimal, anti-commuting and global in nature. The finite nature of this parameter was first discussed in the work of Igarashi and Kubo\cite{IK}. Later Joglekar and Mandal \cite{JM1} studied it in detail in their insightful work, leading to finite field-dependent BRST (FFBRST) transformations, which was further developed by Joglekar etall \cite{BJ,JAM1,JAM2,JM,UM}. Since then, it has found applications in various physical models such as in the references\cite{Mandal:2022zil,Upadhyay:2010ww, DPM,URM,PM1,PM2,MPT,PT,LL}.

In this section, we will show how the Greens functions corresponding to two previous gauge-fixed quantum actions $S_u = S_{SZ} + S_{ug} $ and $S_{tt}$ in equation (\ref{Stt}) can be connected in terms of a FFBRST transformation \cite{BJ}. Following the original ideas of Joglekar and Mandal \cite{JM1}, we start by generalizing BRST parameter as
\begin{equation}\label{FBRST}
\varphi \longrightarrow \varphi+(s\varphi) \varTheta'[\varphi_\kappa]d\kappa
\end{equation}
where $\varTheta'[\varphi_\kappa]d\kappa$ plays the role of an infinitesimal, anticommuting field-dependent quantity, expressed in terms of a continuous parameter $\kappa$.  The infinitesimal character of the transformation (\ref{FBRST}) is captured by the differential $d\kappa$, while the Grassmann function $ \varTheta'[\varphi_\kappa] $ guarantees the dependence on field.  The finiteness of an actual FFBRST transformation can then be achieved by integrating in $\kappa$ from $0$ to $1$ to produce
\begin{equation}\label{FFBRST}
\varphi_{\kappa=1}=\varphi_{\kappa=0}+(s\varphi)\varTheta[\varphi]
\,,
\end{equation}
where
\begin{equation}
\varTheta[\varphi]=\varTheta'[\varphi]\frac{\exp f[\varphi] - 1}{f[\varphi]}
\end{equation}
corresponds to the finite field-dependent parameter with $f[\varphi]$ given by 
\begin{equation}
f[\varphi] = \sum_\varphi \int d^{26}x \frac{\delta \varTheta'[\varphi] }{\delta \varphi } (s\varphi)
\label{FBR}
\end{equation}
summing over all fields $\varphi$.
It has been found that a FFBRST transformation of the form (\ref{FBRST}) leads to a nontrivial Jacobian in the integration functional measure.

\subsection{Improved FFBRST Transformation Connecting Green's function in two different gauges of String Theory}
In this section, we establish a procedure for a FFBRST transformation that transforms the generating functional (Green’s function) in one kind of a gauge choice to the generating functional in another kind of a gauge choice. For this purpose we define the generating functional for Siegel-Zwiebach action from String theory in a transverse-traceless gauge,
\begin{equation}
W_{TT} = \int D\varphi \ e^{iS_{TT}[\varphi]}
\end{equation}
which transforms under a FFBRST transformation $\varphi'(x) = \varphi(x) + s\varphi\varTheta[\varphi]$ defined in \cite{JM1} as follows:
\begin{equation}
W_{FP} = \int D\varphi' \ e^{iS_{FP}[\varphi']} = W_{TT}
\end{equation}
Now, we will implement this transformation to connect
the Green’s functions in the two gauges for the SZ action in String theory. According to the standard procedure, n-point Green’s functions in a Fierz-Pauli gauge under the FFBRST transformation modifies as
\begin{eqnarray}
G^{FP}_{i_1,...,i_n} &=& \int D\varphi' \Pi_{r=1}^{n}\varphi'_{i_r} e^{iS_{FP}[\varphi']} \nonumber\\
&=&\int D\varphi' \Pi_{r=1}^{n}(\varphi_{i_r}(x) + s_{i_r;BRST}\varphi\varTheta[\varphi]) e^{iS_{TT}[\varphi']} \nonumber\\
&=& G^{TT}_{i_1,...,i_n} + \Delta G^{TT}_{i_1,...,i_n},
\end{eqnarray}
where $\Delta G^{TT}_{i_1,...,i_n}$ refers to n-point Green's functions in two sets of gauges. This may involve additional vertices corresponding to insertions of operators $s_{BRST}\varphi$. But it seems technically incorrect due to following reasons.

A priori, it is not obvious that if condition for replacing the Jacobian for $e^{iS_1}$ \cite{JM1,BJ} holds for the given theory; then an operator modified to include an arbitrary operator $O[\varphi]$ of type 
\begin{eqnarray}
&&\int D\varphi O[\varphi] \left[\frac{1}{J(\kappa)}\frac{d J(\kappa)}{d \kappa}-i\frac{dS_1[\varphi(x,k),k]}{d\kappa}\right] \nonumber\\&&e^{i \displaystyle (S_{FP}[\varphi] + S_1[\varphi,k])} = 0 
\label{ffbc}
\end{eqnarray}
would also hold. Of course it doesn't hold in general for reason discussed in \cite{JM1,BJ}. For these reasons, to connect the Green's functions for the two type of gauges we need a detailed treatment of the FFBRST transformation. 

We begin with the Green's function in FP gauge defined as
\begin{equation}
G = \int D\varphi' O[\varphi']\ e^{iS_{FP}[\varphi']}
\label{AFP}
\end{equation}
where $O[\varphi']$ is an arbitrary operator. Hence, (\ref{AFP}) covers both the arbitrary operator Green’s functions and the arbitrary ordinary Green’s functions. We want to express the Green’s function (G) of SZ action from string theory entirely in terms of the transverse-traceless gauge
Green’s functions. So we define
\begin{equation}
G(\kappa) = \int D\varphi O[\varphi(\kappa),\kappa]\ e^{i(S_{TT}[\varphi] + S_1[\varphi,\kappa])}
\label{Gk}
\end{equation}
where the form of operator $O[\varphi(\kappa),\kappa ]$ demands
\begin{equation}
\frac{dG}{d\kappa} = 0
\label{dGdk}
\end{equation}
Under a FFBRST transformation $(\kappa = 1)$, it reflects that
\begin{equation}
G(1) = \int D\varphi' O[\varphi, 1 ]\ e^{iS_{FP}[\varphi']}
\label{G1}
\end{equation}
which coincides with (\ref{AFP}), whereas at $\kappa = 0$ this reads
\begin{equation}
G(0) = \int D\varphi \tilde O[\varphi, 0 ]\ e^{iS_{TT}[\varphi]},
\label{G0}
\end{equation}
and is numerically equal to (\ref{G1}). Now, we need to determine the form of $O[\varphi(\kappa), \kappa]$ in (\ref{Gk}) so that the condition (\ref{dGdk}) is satisfied. For this purpose, we perform the field transformation from $\varphi(\kappa)$ to $\varphi(\kappa +d\kappa)$ through an infinitesimal field-dependent BRST transformation defined in (\ref{FBRST}), which leads to
\begin{eqnarray}
G(\kappa) &=& \int D\varphi(\kappa + d\kappa)\frac{J(\kappa + d\kappa) }{J(\kappa)} \nonumber\\&\times& \biggl(O[\varphi(\kappa+d\kappa),\kappa+d\kappa]-s\varphi\varTheta'\frac{\delta O}{\delta\varphi}d\kappa + \frac{\partial O}{\partial\kappa}d\kappa\biggl)\nonumber\\
&\times&\biggl(1-i\frac{dS_1}{d\kappa}d\kappa\biggl) e^{i(S_{TT}[\varphi(\kappa + d\kappa)] + S_1[\varphi(\kappa + d\kappa),\kappa+d\kappa])}\nonumber\\
&=& \int D\varphi(\kappa + d\kappa)\biggl(1+\frac{1}{J}\frac{d J}{d(\kappa)}d\kappa\biggl) \nonumber\\&\times& \biggl(O[\varphi(\kappa+d\kappa),\kappa+d\kappa]-s\varphi\varTheta'\frac{\delta O}{\delta\varphi}d\kappa + \frac{\partial O}{\partial\kappa}d\kappa\biggl)\nonumber\\
&\times&\biggl(1-i\frac{dS_1}{d\kappa}d\kappa\biggl) e^{i(S_{TT}[\varphi(\kappa + d\kappa)] + S_1[\varphi(\kappa + d\kappa),\kappa+d\kappa])}\nonumber\\
&=& G(\kappa + d\kappa)
\end{eqnarray}
if and only if  
\begin{eqnarray}
&&\int D\varphi(\kappa ) \biggl(\biggl[\frac{1}{J}\frac{d J}{d(\kappa)}-i\frac{dS_1}{d\kappa}\biggl]O[\varphi(\kappa),\kappa]-s\varphi\varTheta'\frac{\delta O}{\delta\varphi}d\kappa \nonumber\\&& + \frac{\partial O}{\partial\kappa}d\kappa\biggl) e^{i(S_{TT}[\varphi(\kappa)] + S_1[\varphi(\kappa),\kappa])} = 0
\label{CFBRST}
\end{eqnarray}
So we get precisely the correct expression (\ref{CFBRST})for replacing the Jacobian of the path integral measure in the Green’s function of SZ action from String theory as $e^{iS_1}$ instead of the incorrect one in (\ref{ffbc}).

Using the information above, the required condition for the $\kappa$-independence of G is
\begin{eqnarray}
&&\int D\varphi(\kappa )e^{i(S_{TT}[\varphi(\kappa)] + S_1[\varphi(\kappa),\kappa])} \biggl( \frac{\partial O}{\partial\kappa} + \int (\partial_\mu c_\nu + \partial_\nu c_\mu \nonumber\\&&- \frac{m}{2} \eta_{\mu\nu} c)\varTheta'\frac{\delta O}{\delta h_{\mu\nu}} + \int b_\mu\varTheta'\frac{\delta O}{\delta {\bar c_\mu}} + \int b\varTheta'\frac{\delta O}{\delta {\bar c}}+ \int \biggl[\partial_\mu c \nonumber\\&&+ m c_\mu\biggl]\varTheta' \frac{\delta O}{\delta B_\mu} + \int \biggl[-\partial_\mu \tilde c^\mu + \frac{3}{2} m \tilde c\biggl]\varTheta' \frac{\delta O}{\delta \theta}\biggl)  = 0  
\label{OPZ}
\end{eqnarray} 
Now, we construct the operator O to satisfy
\begin{eqnarray}
&&\frac{\partial O}{\partial\kappa} + \int (\partial_\mu c_\nu + \partial_\nu c_\mu - \frac{m}{2} \eta_{\mu\nu} c)\varTheta'\frac{\delta O}{\delta h_{\mu\nu}} + \int b_\mu\varTheta'\frac{\delta O}{\delta {\bar c_\mu}} \nonumber\\&&+ \int b\varTheta'\frac{\delta O}{\delta {\bar c}}+ \int \biggl[\partial_\mu c + m c_\mu\biggl]\varTheta' \frac{\delta O}{\delta B_\mu} + \int \biggl[-\partial_\mu \tilde c^\mu \nonumber\\&&+ \frac{3}{2} m \tilde c\biggl]\varTheta' \frac{\delta O}{\delta \theta} = 0
\label{Op}
\end{eqnarray} 
Then condition (\ref{OPZ}) automatically is satisfied. Now, we consider a new set of fields $(\tilde h_{\mu\nu},\tilde B_\mu, \tilde \theta, \tilde {\bar c}, \tilde {\bar c}_\mu, \tilde c_\mu, \tilde c, \tilde b_\mu, \tilde b)$ having the following infinitesimal field-dependent BRST transformation:
\begin{eqnarray}
\frac{\delta \tilde h_{\mu\nu}}{\delta \kappa} &=&  (\partial_\mu \tilde c_\nu + \partial_\nu \tilde c_\mu - \frac{m}{2}\tilde \eta_{\mu\nu}\tilde c)\varTheta'[\tilde\varphi],\nonumber\\
\frac{\delta \tilde {\bar c}_\mu}{\delta \kappa} &=&  \tilde b_\mu\varTheta'[\tilde\varphi],\nonumber\\
\frac{\delta \tilde {\bar c}}{\delta \kappa} &=&  \tilde b\varTheta'[\tilde\varphi],\nonumber\\
\frac{\delta \tilde B_\mu}{\delta \kappa} &=& (\partial_\mu \tilde c + m \tilde c_\mu)\varTheta'[\tilde\varphi],\nonumber\\
\frac{\delta \tilde \vartheta}{\delta \kappa} &=& (-\partial_\mu \tilde c^\mu + \frac{3}{2} m \tilde c)\varTheta'[\tilde\varphi],\nonumber\\
\frac{\delta \tilde {c}_\mu}{\delta \kappa} &=& \frac{\delta \tilde {c}}{\delta \kappa} = \frac{\delta \tilde {b}_\mu}{\delta \kappa} = \frac{\delta \tilde {b}}{\delta \kappa} = 0
\label{IFFBRST}
\end{eqnarray} 
These new fields satisfy following boundary condition: $\tilde\varphi(1) = \phi(1)$. The condition (\ref{Op}) for $O[\tilde\varphi(\kappa), \kappa]$ instead $O[\varphi(\kappa), \kappa]$ reads
\begin{equation}
\frac{dO[\tilde\varphi(\kappa), \kappa]}{d\kappa} = 0
\end{equation}
Now utilizing $O[\tilde\varphi(1), 1] = O[\varphi(1), 1] = O[\varphi']$ we obtain 
\begin{equation}
O[\tilde\varphi(\kappa), \kappa] = O[\varphi'],
\end{equation}
which tells us how the operator $O[\varphi(\kappa), \kappa] $evolves. To derive the FFBRST transformation corresponding to (\ref{IFFBRST}), we first define the modification in f of (\ref{FBR}) as follows:
\begin{equation}
f[\tilde\varphi,\kappa] = f_1[\tilde\varphi] + \kappa f_2[\tilde\varphi].
\end{equation}
Performing an integration of eqn(\ref{FBRST}) from 0 to $\kappa$, we have
\begin{equation}
\varTheta'[\tilde\varphi(\kappa)] = \varTheta[\varphi]\exp\biggl(\kappa f_1[\varphi] + \frac{\kappa^2}{2}f_2[\varphi]\biggl) .
\end{equation}
Similarly, integrating (\ref{IFFBRST}) we get the FFBRST transformation, written compactly as
\begin{eqnarray}
\varphi' &=& \varphi + \biggl[(\tilde\delta_1[\varphi]+\tilde\delta_2[\varphi])\int d\kappa \exp \nonumber\\&& \biggl( \kappa f_1[\varphi] + \frac{\kappa^2}{2}f_2[\varphi]\biggl)\biggl]\varTheta'[\varphi]\nonumber\\ 
&=& \varphi + \delta \varphi[\varphi]
\label{NFBRST}
\end{eqnarray}
Now we apply the FFBRST transformation (\ref{NFBRST}) on the Green’s function in Fierz-Pauli gauge(\ref{AFP})
\begin{eqnarray}
&&G = \int D\varphi'O[\varphi']e^{iS_{FP}[\varphi']},\nonumber\\
&&= \int D\varphi O[\varphi + \delta \varphi[\varphi]]e^{iS_{TT}[\varphi]}\nonumber\\ &&+ \int D\varphi \biggl[(\tilde\delta_1[\varphi]+\tilde\delta_2[\varphi])\int d\kappa \exp \biggl( \kappa f_1[\varphi] + \frac{\kappa^2}{2}f_2[\varphi]\biggl)\biggl]\nonumber\\&&\varTheta'[\varphi]\frac{\delta O}{\delta \varphi}e^{iS_{TT}[\varphi]}
\end{eqnarray}
Further, it can be written by 
\begin{eqnarray}
&&\langle O \rangle_{FP} = \langle O \rangle_{TT} \nonumber\\&&+ \int_0^1 d\kappa \int D\varphi[(\tilde\delta_1[\varphi]+\tilde\delta_2[\varphi])]\varTheta'[\varphi]\frac{\delta O}{\delta \varphi}e^{iS_M}
\label{FFGr}
\end{eqnarray} 
where $iS_M = iS_{TT} + \kappa f_1[\varphi] + \frac{\kappa^2}{2}f_2[\varphi]$. In this way, we establish the connection between the Green’s function of two gauges in Seigel-Zweibach action of string theory.

\subsection{Relation between Green’s functions for Fierz-Pauli and Transverse-Traceless gauge}
In the present case finite BRST parameter can be written as
\begin{equation}
\varTheta'[\tilde\varphi(\kappa)] = \varTheta[\varphi]\exp\biggl(\kappa f_1[\varphi] + \frac{\kappa^2}{2}f_2[\varphi]\biggl) .
\end{equation}
Here $f_1[\phi]$ and $f_2[\phi]$ are defined as
\begin{eqnarray}
&&f_1[\varphi] \equiv i \int d^{26} x \bigg[b_\mu \bigg(\frac{\xi b^\mu}{2} + \partial^\nu \eta^{\mu\alpha}h_{\alpha\nu} - B^\mu\bigg) + b \bigg(\frac{\zeta b}{2m^2} \nonumber\\&&+ \frac{1}{2}\eta^{\mu\nu}h_{\mu\nu} - \theta\bigg) + \bar c_\mu \bigg(\eta^{\mu\alpha}\partial^\nu(\partial_\alpha c_\nu + \partial_\nu c_\alpha - \frac{m}{2}\Gamma_{\alpha\nu}c) - \partial_\mu c - m c_\mu \bigg)\nonumber\\&&+ \bar c\bigg(\frac{1}{2}\eta^{\mu\nu}(\partial_\mu c_\nu + \partial_\nu c_\mu - \frac{m}{2}\Gamma_{\mu\nu}c) - \partial_\mu c^\mu + \frac{3}{2}m c\bigg)\bigg]
\label{f1}
\end{eqnarray}
and 
\begin{eqnarray}
&&f_2[\varphi] \equiv -i \int d^{26} x \bigg[ \bigg(\frac{\xi b^\mu}{2} + \partial^\nu h^\mu_\nu - B^\mu \nonumber\\&&+ \frac{\zeta b}{2m^2} + \frac{h}{2} - \theta\bigg)^2 \bigg]
\label{f2}
\end{eqnarray}
Now the Greens's function in two gauges can be connected using eqn({\ref{FFGr}}).
where
\begin{eqnarray}
\langle O \rangle_{FP} = iG^{FP}_{\mu\nu\alpha\beta}
\end{eqnarray} 
and 
\begin{eqnarray}
\langle O \rangle_{TT} = iG^{TT}_{\mu\nu\alpha\beta}
\end{eqnarray} 
It can be written explicitly as
\begin{eqnarray}
&&iG^{FP}_{\mu\nu\alpha\beta} = iG^{TT}_{\mu\nu\alpha\beta} + i\int_0^1 d\kappa \int D\varphi[(\tilde\delta_1[\varphi]+\tilde\delta_2[\varphi])] \nonumber\\&&\int d^{26} x \bigg[ {\bar c}_\mu\bigg(\frac{\xi b^\mu}{2} + \partial^\nu h^\mu_\nu - B^\mu\bigg) + \bar c\bigg(\frac{\zeta b}{2m^2} \nonumber\\&&+ \frac{h}{2} - \theta\bigg) \bigg]\frac{\delta O}{\delta \varphi}e^{iS_M}
\label{GFPTT}
\end{eqnarray} 
Propagator $O$ is defined in present case as
\begin{eqnarray}
&&O^{FP}_{\mu\nu\alpha\beta} = -\frac{p^2 + m^2}{4}\bigg(\eta_{\mu\alpha}\eta_{\nu\beta} + \eta_{\nu\alpha}\eta_{\mu\beta} - 2 \eta_{\mu\nu}\eta_{\alpha\beta}\bigg)\nonumber\\&&-\frac{1}{2}\bigg(\eta_{\mu\nu}p_{\alpha}p_{\beta} + \eta_{\alpha\beta}p_{\mu}p_{\nu}\bigg) + \frac{1}{4}\bigg(\eta_{\mu\alpha}p_{\nu}p_{\beta} + \eta_{\mu\beta}p_{\nu}p_{\alpha}\nonumber\\&& + \eta_{\nu\alpha\beta}p_{\mu}p_{\beta} + \eta_{\nu\beta}p_{\mu}p_{\alpha}\bigg)
\end{eqnarray}
Where $S_M$ is defined above and $f_1$ and $f_2$ are defined in eqns(\ref{f1}) and (\ref{f2}).

Further simplification of eqn(\ref{GFPTT}) will lead to terms dependent on $\kappa$ \cite{BJ,JAM1,JAM2} which will remain finite in the limit $m\rightarrow 0$ making the Green's function $G^{FP}_{\mu\nu\alpha\beta}$ finite in the massless limit resolving the problem of divergence of propagator in Fierz-Pauli Gauge. 

\section{Conclusion}
In the present work continuing the previous work \cite{PT} we have provided a correct prescription to the divergence problem of the massive symmetric rank two tensor field propagator in the  string critical dimension. We have shown how this problem can be rigorously addressed using the FFBRST transformation approach by explicitly connecting the Green's functions of two set of generating functionals. Our method has proved to lead to a significant result for the small mass graviton propagator (Green's function) in the linear form of gravity which has been found to be finite in the massless limit in traceless transverse gauge. By means of the Improved FFBRST technique, we have connected Green's function of a Fierz-Pauli type gauge to of transverse-traceless gauge, arising from the open string action. The improved FFBRST transformations have generated a
nontrivial Jacobian in the quantum generating functional path-integral integration measure which has been responsible for the proper connection of the Greens's functions of aforementioned gauges. The Green’s functions in Fierz-Pauli gauge in the theory of string theory is expressed as a series in terms of those in transverse-traceless gauge leading towards solution to problem of divergence of propagator of SZ action in FP gauge in the massless limit.

\end{document}